\documentclass[11pt]{article}
\usepackage{amsmath,amssymb}
\usepackage{graphicx}
\begin{document}

\noindent
\centerline{ \LARGE \bf Frequency stabilization by synchronization}
\centerline{ \LARGE \bf of Duffing oscillators}

\vspace{10 pt}

\centerline{\large Dami\'an H. Zanette}

\vspace{10 pt}
\noindent
{\it Centro At\'omico Bariloche and Instituto Balseiro (Comisi\'on Nacional de Energ\'{\i}a At\'omica, Universidad Nacional de Cuyo), Consejo Nacional de Investigaciones Cient\'{\i}ficas y T\'ecnicas. 8400 San Carlos de Bariloche, R\'{\i}o Negro, Argentina}


\vspace{10 pt}

{\bf Abstract} -- We present analytical and numerical results on the joint dynamics of two coupled Duffing oscillators with nonlinearity of opposite signs (hardening and softening). In particular, we focus on the existence and stability of synchronized oscillations where the frequency is independent of the amplitude. In this regime, the amplitude--frequency interdependence (a--f effect) ---a noxious consequence of nonlinearity, which jeopardizes the use of micromechanical oscillators in the design of time--keeping devices--- is suppressed. By means of a multiple time scale formulation, we find approximate conditions under which frequency stabilization is achieved, characterize the stability of the resulting oscillations, and compare with numerical solutions to the equations of motion.


\vspace{10 pt}

{\bf Introduction} -- The practical problem of transforming a mechanical oscillator into a clock ---brilliantly solved for the first time by unknown engineers more than a millennium ago, probably in China \cite{China}--- requires achieving two key goals. First, the oscillator must sustain periodic motion upon stationary energy supply from outside the system. Second, the oscillation frequency must be autonomously generated by the system itself, as an emergent dynamical property, without need of any external periodic input. The escapement mechanism of pendulum clocks and mainspring watches is an ingenious answer to these requirements \cite{escap}. 

\begin{figure}
\centerline{\includegraphics[width=.8\textwidth]{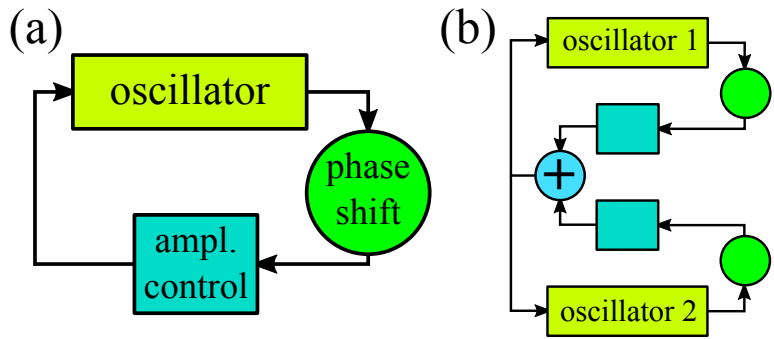}}
\caption{(a) Schematic representation of a self--sustained oscillator. The signal read from the oscillator is conditioned by phase shifting and amplitude controlling before being reinjected as a self--sustaining force. (b) Two self--sustained oscillators coupled through their conditioned signals, as in eq. (\ref{ensemble}).} \label{fig1}
\end{figure}

In modern everyday clocks, the same basic principle is implemented by using, as oscillator, a piezoelectric (usually, quartz) crystal. The crystal vibrations generate an electric signal which is input to an electronic circuit. This  signal is conditioned by shifting its phase by a prescribed amount and fixing its amplitude to a given value, and then reinjected as an electric force acting on the crystal, as schematized in fig.~\ref{fig1}(a) \cite{Yurke}. The crystal, in turn, responds to this driving force as a mechanical resonator. As the result of this feedback loop, a self--sustained oscillation is established, whose amplitude and frequency are determined by the mechanical properties of the crystal and the parameters of signal conditioning (see next section). The only input from outside the system is the power which feeds the conditioning circuit, typically supplied by a DC battery. 

Quartz crystals, however, are difficult to miniaturize, which limits their application to time--keeping components in microelectronic devices. Tiny vibrating silica bars, which can be readily integrated to circuits during fabrication and can be actuated by relatively weak electric fields, have been proposed as a substitute \cite{microos,review}. To overcome the effects of thermal and electric noise, these micromechanical oscillators must vibrate at large amplitudes, well within a dynamical regime where nonlinear effects cannot be neglected \cite{microos1}. A direct consequence of nonlinearity on oscillatory motion is that  amplitude and frequency become interdependent quantities (amplitude--frequency, or a--f, effect). In particular, a variation of the amplitude ---due, for instance, to fluctuations in the self--sustaining force--- brings about a change in the frequency, which is obviously undesirable in any application to time--keeping devices. A crucial problem in these applications is therefore how to suppress or at least minimize the a--f effect, by stabilizing the oscillation frequency against amplitude variations.  

Recently, a mechanism of frequency stabilization in micromechanical oscillators has been proposed and demonstrated experimentally, relying on the mutual resonance of different oscillation modes \cite{Nat,EPJB}. Here, we explore a different theoretical approach to the same problem, based on the joint dynamics of two coupled oscillators with nonlinearity of opposite signs. The underlying idea is that frequency deviations in opposite directions can compensate each other if the oscillators reach mutually synchronized motion. In the next section, we summarize previous results on the self--sustained Duffing oscillator, which provides a standard model for nonlinear micromechanical oscillators. Then, we find the conditions of frequency stabilization for two coupled Duffing oscillators with opposite nonlinearity, and characterize their joint dynamics under such conditions, analyzing the existence and stability of synchronized motion. Results for the limit of small damping, which is particularly relevant to microtechnologies, are  presented in detail. Our conclusions have not only potential relevance to technological applications, but are also significant to the field of nonlinear oscillating systems. 

\vspace{10 pt}

{\bf The self--sustained Duffing oscillator} -- The micromechanical oscillators proposed for miniaturized time--keeping devices ---and already employed in nanomechanical sensors--- are structurally designed as beams clamped at both ends (clamped--clamped, or c--c, beams) \cite{microos1}. The leading nonlinear contribution to their dynamics is a cubic restoring force, which adds to the ordinary linear elastic force \cite{tufi}. In its main oscillation mode, thus, the vibrating self--sustained c--c beam is well described by the Duffing equation for a coordinate $x(t)$ which represents the displacement from equilibrium \cite{Nayfeh}. Normalizing by the effective mass, the equation reads \cite{Nat,EPJB,EPJST}
\begin{equation} \label{Duff}
\ddot x+ \mu \dot x +\omega^2 x+\beta x^3 =f_0 \cos (\phi+\phi_0),
\end{equation}
where $\mu$ is the damping coefficient per unit mass, $\omega$ is the frequency of the undamped ($\mu=0$), linear ($\beta=0$), unforced ($f_0=0$) oscillator, and $\beta$ weights the nonlinear force per unit mass. The right--hand side of eq.~(\ref{Duff}) stands for the self--sustaining force. Here, $\phi$ is the phase of the oscillation. In harmonic motion,  this phase is defined through the relation $x (t)\propto \cos \phi (t) $ (see below).  The phase shift $\phi_0$ and the amplitude of the force per unit mass $f_0$ are fixed by the process of signal conditioning.  In the following, we focus on the choice $\phi_0=\pi/2$, for which the self--sustaining force and the oscillation velocity $\dot x(t) \propto -\sin \phi (t)$ are in--phase, and the resonant response of the oscillator is therefore maximal. In this situation, in fact, $\cos (\phi+\phi_0) =-\sin \phi $.

A standard approximation to deal with the Duffing equation is the multiple scale method \cite{Nayfeh}. To the lowest significant order,  where higher--harmonic contributions in the nonlinear term are neglected, the method assumes an oscillating solution with slowly changing amplitude, and whose phase varies with the natural frequency $\omega$ plus a slowly changing phase shift, namely, $x(t) = A(\epsilon t) \cos [\omega t +\psi (\epsilon t)]$. The small quantity  $\epsilon$ defines a slow time scale $\epsilon t$, and acts as the perturbative parameter in the approximation. The above solution works if, in eq.~(\ref{Duff}), all the forces acting on the oscillator are of order $\epsilon^1$, except for the linear elastic force, of order $\epsilon^0$. The method yields equations of motion for $ A(\epsilon t)$ and $\psi(\epsilon t)$, thus making it possible to study stationary solutions and their stability. 

Within this approach, it can be shown that the solution to eq.~(\ref{Duff}) attains asymptotic oscillations whose amplitude and frequency are 
\begin{equation} \label{Omega}
A=\frac{f_0}{\mu \omega}, \ \ \ \ \  \Omega = \omega \left( 1 + \frac {3 \beta f_0^2}{8\mu^2\omega^4} \right).
\end{equation}
The dependence of $\Omega$ on $f_0$ ---or, equivalently, the interdependence of $\Omega$ and $A$--- becomes significant when the nonlinear force overcomes the damping force, $\beta A^3 \gtrsim \mu \Omega A$. This is the undesirable a--f effect quoted above \cite{EPJST}.

\vspace{10 pt}

{\bf Synchronized oscillators and frequency stabilization}  -- Equation (\ref{Omega}) makes it clear that the nonlinear correction to the oscillation frequency depends on the sign of the cubic coefficient $\beta$. Respectively, for hardening and softening nonlinearity (positive and negative $\beta$), $\Omega$ becomes increasingly larger and smaller than the natural frequency $\omega$ as the amplitude grows.  Just like for a vibrating string fixed at its two ends \cite{tufi}, cubic nonlinearity in a c--c beam is hardening ($\beta >0$) and, therefore, its frequency grows with the amplitude \cite{Nat}. It has been shown, however, that other kinds of microoscillators ---in particular, mechanical elements vibrating in torsional modes--- exhibit the opposite behavior \cite{Nat,tors}, and can be described by the Duffing equation with negative $\beta$. The question thus arises whether this contrasting response may be exploited to counteract the a--f effect. Specifically, we inquire if the interaction between two oscillators with nonlinearity of opposite signs may help to neutralize the a--f effect by the mutual compensation of their individual behavior.

To explore this problem, we first define a specific form of coupling between two self--sustained Duffing oscillators. It consists in  replacing the self--sustaining force by a sum of the forces acting on each oscillator. Namely, we propose
\begin{equation} \label{ensemble}
\ddot x_j+ \mu_j \dot x_j +\omega_j^2 x_j +\beta_j x_j^3= -  f_i ( \sin \phi_1 +\sin \phi_2) ,  
\end{equation}
where the index $j=1,2$ identifies the quantities associated to each individual oscillator. The mutual interaction introduced by the sum of driving forces, which belongs to a class of mean--field coupling extensively studied in the literature \cite{sync}, has the advantage that it can be straightforwardly implemented in experiments by means of a simple summing electric circuit, as schematized in fig.~\ref{fig1}(b). The factors $f_i$ have been introduced to take into account that, experimentally, the same coupling force may affect each oscillator with different strength, depending on the transducer mechanism used in each case to reinject the driving signal. However, these factors can be rescaled by a convenient choice of the units in which the coordinates $x_j$ are measured, and/or by redefining the other coefficients in eq.~(\ref{ensemble}). Therefore, without generality loss, we take $f_1=f_2\equiv f_0$.

Equations (\ref{ensemble}) can be dealt with by the multiple scale method quoted in the preceding section \cite{Nayfeh}. We propose solutions of the form $x_j   = A_j \cos \phi_j \equiv A_j \cos (\omega_j t+\psi_j)$, where $A_j$ and $\psi_j$ vary slowly with time as compared with the oscillations of frequency $\omega_j$. Moreover, we assume that the two natural frequencies $\omega_1$ and $\omega_2$ differ by a quantity of the same order as the perturbative parameter $\epsilon$. 

Stationary solutions to the equations of motion for amplitudes and phases show that the two oscillators can attain synchronized motion, with a common oscillation frequency $\Omega$. The stationary amplitudes $A_1$ and $A_2$, the stationary phase difference $\Delta=\phi_2-\phi_1$, and the frequency $\Omega$ satisfy the four algebraic equations
\begin{equation} \label{st1}
2\omega_1 \nu_1 A_1 -\frac{3}{4}  \beta_1 A_1^3= -2\omega_2 \nu_2 A_2 +\frac{3}{4} \beta_2 A_2^3=f_0 \sin \Delta,
\end{equation}
with  $\nu_j =\Omega-\omega_j$, and
\begin{equation}  \label{st2}
\mu_1 \omega_1 A_1=\mu_2 \omega_2 A_2=f_0 (1+\cos \Delta).
\end{equation}

Equations (\ref{st2}) suggest that it is useful to rescale the stationary amplitudes by introducing a new unknown $a$ such that $A_j = a /\mu_j \omega_j$, which reduces the two equations to the single identity $a =f_0(1+\cos \Delta)$. Meanwhile, the first of eqs.~(\ref{st1}) implies
\begin{equation} \label{ec1a}
\frac{\Omega -\omega_1}{\mu_1} +\frac{\Omega -\omega_2}{\mu_2}    = z\,  a^2,
\end{equation}
with 
\begin{equation} \label{z}
z= \frac{3}{8}  \left( \frac{ \beta_1}{\mu_1^3 \omega_1^3} + \frac{\beta_2}{\mu_2^3 \omega_2^3}\right).
\end{equation}
Eliminating $\Delta$ from the above relations yields high--degree polynomial equations for $\Omega$ and $a$, which cannot be solved analytically but admit standard numerical treatment. Equation (\ref{ec1a}), however, makes it clear that the problem simplifies drastically if 
\begin{equation} \label{cond}
z =0 .
\end{equation}
In fact, this condition decouples the interdependence between frequency and amplitude, yielding
\begin{equation} \label{stabfrec}
\Omega_{z=0}= \frac{\mu_2 \omega_1 + \mu_1 \omega_2}{\mu_1+\mu_2},
\end{equation}
which does not depend on $f_0$ or $a$. If condition (\ref{cond}) is fulfilled, the frequency is thus {\em stabilized} by mutual synchronization of the oscillators and the a--f effect is suppressed.

Note that the stabilization condition (\ref{cond}) necessarily requires that the nonlinearity in the two oscillators has opposite signs. In the following, without generality loss, we take $\beta_2<0<\beta_1$. Under frequency stabilization, the synchronization frequency $\Omega$ always lies between the frequencies $\omega_1$ and $\omega_2$, and is closer to the frequency of the oscillator with the smaller damping coefficient. In view of the behavior of the self--sustained Duffing oscillator described in the preceding section, because of the sign of the respective cubic coefficients, coupling makes the frequency of oscillators 1 and 2 respectively grow and decrease. Then, synchronization will effectively be possible if $\omega_1 < \omega_2$. As we show in the next section, this requisite reflects into the  stability condition for synchronized motion.

\begin{figure}
\centerline{\includegraphics[width=.8\textwidth]{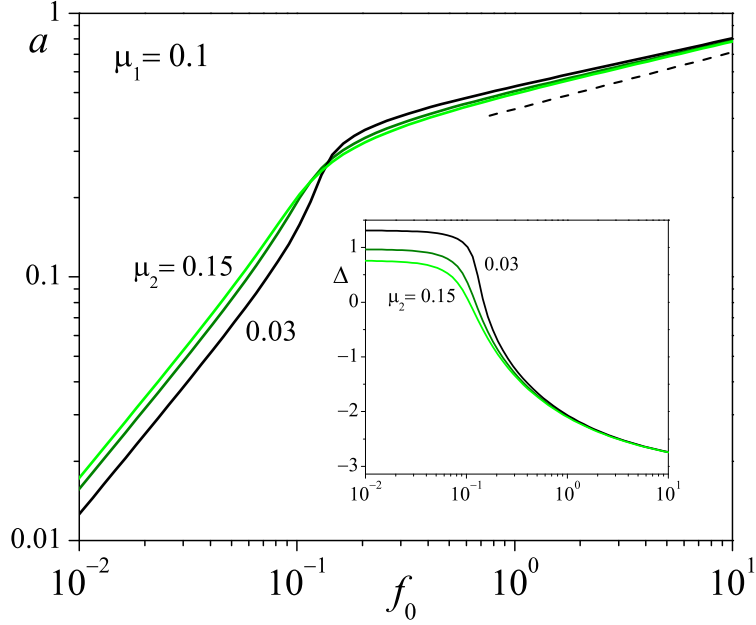}}
\caption{The rescaled stationary amplitude $a=\mu_j \omega_j A_j$ as a function of $f_0$ under frequency stabilization conditions, for $\omega_1=0.95$ and $\omega_2=1$, $\mu_1=0.1$, and $\mu_2=0.03$, $0.09$ (not labeled), and $0.15$. The nonlinear coefficient of the oscillator with hardening nonlinearity is  $\beta_1=0.01$, while $\beta_2$ is chosen in such a way that eq.~(\ref{cond}) is fulfilled  ($\beta_2<0$). The slope of the dashed straight line is $1/5$. The inset shows the corresponding dependence of the relative phase shift $\Delta$.} \label{fig2}
\end{figure}

In contrast with the synchronization frequency $\Omega$, the amplitudes $A_j$ preserve a nontrivial dependence on the coupling force amplitude $f_0$. To illustrate this fact with a concrete example, we choose $\omega_1=0.95$, $\omega_2=1$, $\mu_1=0.1$, and consider various values of $\mu_2$ (see discussion in the final section). Moreover, we take $\beta_1=0.01$, tuning $\beta_2$ in such a way that the frequency stabilization condition (\ref{cond}) holds.  From the above relations, we numerically calculate the rescaled stationary amplitude $a= \mu_j \omega_j A_j$ and the relative phase shift $\Delta$ as functions of $f_0$. Results are shown in fig.~\ref{fig2} and its inset, for $\mu_2=0.03$, $0.09$, and $0.15$. 

The limits of small and large coupling force can be worked out analytically. In the former, the oscillators behave linearly and, consequently, the amplitudes are proportional to the force, $a\approx [1+4(\omega_2-\omega_1)^2/(\mu_1+\mu_2)^2]^{-1} f_0$. The corresponding value of $\Delta$ is positive. For large $f_0$, on the other hand, the amplitudes exhibit a strongly sublinear growth with the coupling force, $a \approx 2 (\mu_j^6 \omega_j^6/\beta_j^2)^{1/5} f_0^{1/5}$ (by virtue of the stabilization condition, the prefactor in this last expression is the same for $j=1$ and $2$). Meanwhile, $\Delta$ asymptotically approaches $-\pi$ from above. Therefore, there is an intermediate value of  $f_0$ at which $\Delta=0$, where the two coupled oscillators move in phase. At each side of this point, either oscillator precedes its partner along their synchronized motion. Note that $\Delta$ switches its sign in the same zone where $a$ changes its slope.

Stability of stationary motion can be assessed by ordinary linearization of the equations of motion derived from the multiple scale approach. Synchronized oscillations ---both under condition (\ref{cond}) and otherwise--- are stable if the nontrivial eigenvalues of the Jacobian
\begin{equation} \label{mat}
J=\frac{1}{2}
\left( \begin{array}{cccc}
-\mu_1  & 0 & \frac{f_0^s}{\omega_1} & - \frac{f_0^s}{\omega_1} \\ \\ 
0 & -\mu_2& \frac{f_0^s}{\omega_2} & - \frac{f_0^s}{\omega_2} \\ \\
\frac{3 \beta_1 A_1^3 -2f_0^s}{2 \omega_1 A_1^2} & 0 & - \frac{f_0^c}{\omega_1 A_1} & \frac{f_0^c}{\omega_1 A_1} \\ \\
0 & \frac{3 \beta_2 A_2^3 +2f_0^s}{2 \omega_2 A_2^2} &   \frac{f_0^c}{\omega_2 A_2} & -\frac{f_0^c}{\omega_2 A_2}
\end{array} \right) 
\end{equation}
are negative or have negative real parts. Here, $f_0^s = f_0 \sin \Delta$, $f_0^c = f_0 \cos \Delta$,   and the values of $A_1$, $A_2$, and $\Delta$  are the solutions to eqs.~(\ref{st1}) and (\ref{st2}). The fact that the two last columns of $J$ are mutually proportional indicates that one of the  eigenvalues is zero. This is a consequence of the definition of oscillation phases up to an arbitrary additive constant.  As for the other three eigenvalues, it can be seen that ---over a vast  zone of parameter space--- one of them is always negative, and the other two are complex conjugate to each other. Their common real part can change sign, for instance, by varying the coupling force amplitude $f_0$. More specifically, it changes from negative to positive values as $f_0$ is increased. This points to  destabilization of synchronized oscillations through a Hopf bifurcation as the coupling becomes stronger. 

\begin{figure}
\centerline{\includegraphics[width=.8\textwidth]{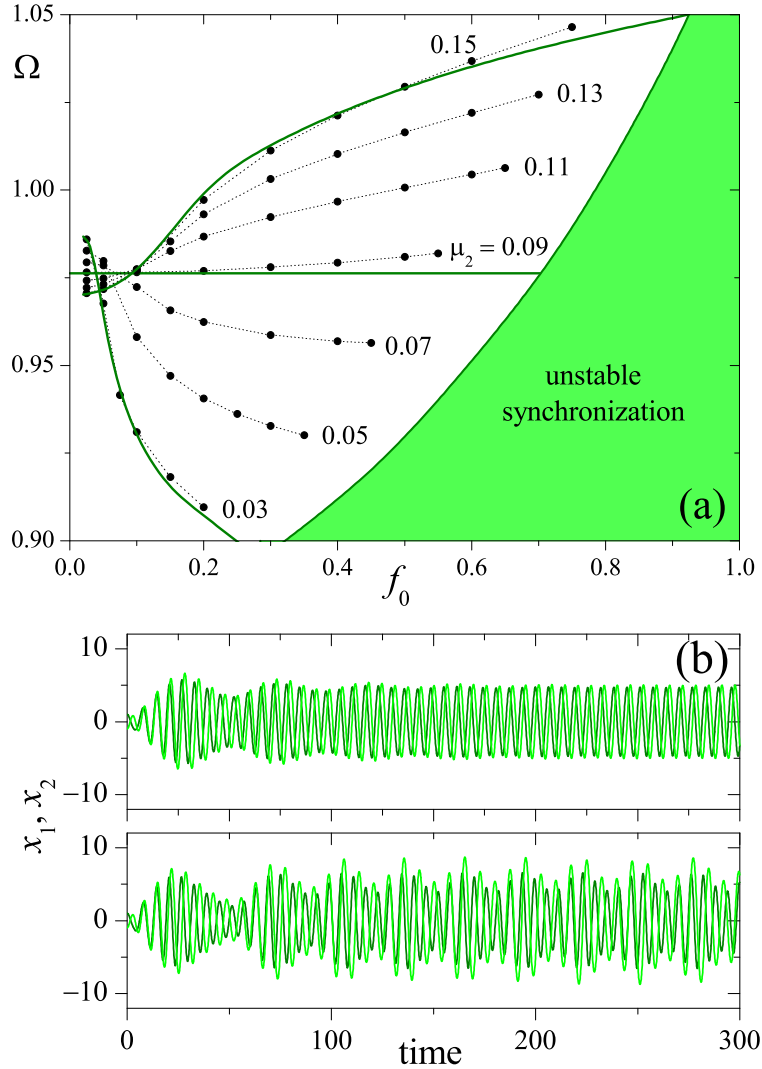}}
\caption{(a) In the shaded region to the right, synchronized oscillations are unstable. To the left, where synchronization is stable, full curves stand for the synchronization frequency $\Omega$ as a function of the combination $f_0$, for $\mu_2=0.03$, $0.09$, and $0.15$, in the multiple scale approximation. The remaining parameters, specified in the main text, have been chosen in such a way that, in the same approximation, the frequency is stabilized for $\mu_2=0.09$.  Full dots joined by dotted lines represent results of the numerical resolution of eqs.~(\ref{ensemble}) for several values of $\mu_2$, indicated by the labels.  (b) Illustration of the joint dynamics of the two Duffing oscillators, as resulting from the numerical resolution of eqs.~(\ref{ensemble}) for the coordinates $x_1(t)$ and $x_2(t)$, in the cases of synchronized (upper panel) and unsynchronized (lower panel) oscillations.} \label{fig3}
\end{figure}

Figure \ref{fig3}(a) shows the regions of stability and instability of synchronized oscillations in the plane spanned by the coupling force amplitude and the synchronization frequency. These results correspond to $\omega_1=0.95$, $\omega_2 =1$, $\mu_1=0.1$, $\beta_1=0.01$, and $\beta_2$ has been chosen in such a way that the frequency stabilization condition is satisfied for $\mu_2=0.09$ (cf. fig.~\ref{fig2}). The boundary of the instability zone has been obtained by detecting the change of sign in the real part of the complex eigenvalues of the Jacobian as $f_0$ is varied, for several values of $\mu_2$, and calculating $\Omega$ for the corresponding set of parameters. Full curves in the plot stand for the synchronization frequency as a function of the coupling force for three selected values of $\mu_2$ ($0.03$, $0.09$, and $0.15$). Because of our choice for $\beta_2$, the value of the frequency for $\mu_2=0.09$, $\Omega \approx 0.976$, does not depend on $f_0$, as it corresponds to the stabilization condition.  

This way of presenting our results is aimed at comparing the analytical approximate formulation provided by the multiple scale method with numerical solutions to the original equations of motion (\ref{ensemble}). Full dots joined by dotted lines in fig.~\ref{fig3}(a) correspond to numerical results obtained by means of a Runge--Kutta scheme for the same parameters quoted above and several values of $\mu_2$, as indicated by the labels. The synchronization frequency was obtained after sufficiently long  transients, by determining the times at which the numerical solutions for $x_1 (t)$ and $x_2 (t)$ crossed zero. Synchronization itself was assessed by direct inspection of the solutions, as illustrated in fig.~\ref{fig3}(b): while in synchronized motion they oscillate quasi--harmonically, both with the same frequency and very well defined amplitudes (upper panel), in unsynchronized motion they display beats and, generally, more incoherent dynamics (lower panel). For each value of $\mu_2$, the set of numerical results in fig.~\ref{fig3}(a) ends at the largest value of $f_0$ for which synchronized oscillations were obtained.

The comparison for $\mu_2=0.03$, $0.09$, and $0.15$, shows that numerical results and the analytical approximation for the synchronization frequency as a function of the coupling force are in very good agreement. The boundaries between the regions of stability and instability obtained from each method also have the same general trend, although the precision of the numerical determination is limited by the resolution with which $f_0$ is varied. As for the suppression of the a--f effect in the numerical results for $\mu_2=0.09$, we see that ---even though frequency stabilization is not perfect--- $\Omega$ varies by considerably less than 1\%  over the interval of coupling forces where synchronization is stable.

\vspace{10 pt}

{\bf The limit of weak  damping}  -- In  usual experimental conditions with micromechanical oscillators, the damping force is normally much weaker than all the other intrinsic mechanical forces acting on the oscillator  \cite{review}. As a consequence, the time scale associated with the effects of damping ---namely, the typical time of energy dissipation--- is substantially longer than any other characteristic time in the dynamics, including not only the natural oscillation period but also any variation in that period caused by nonlinear effects. For c--c beam micromechanical oscillators, the ratio between the damping time and the period of oscillations, which defines its quality factor $Q$, can reach values around $10^4$ to $10^5$ \cite{review,Nat,PRL}. This scale separation justifies the multiple scale method used in the preceding sections and, moreover, allows for further simplifying approximations, as we show in the following. Our main goal is to find explicit analytical expressions for the quantities that characterize stationary synchronized oscillations, as well as their stability. 

In order that the oscillation amplitudes remain finite, the limit of weak damping (small $\mu_j$) must be done together with a limit of weak coupling  (small $f_0$). Physically, this joint limit can be understood by the fact that, in stationary oscillations, the energy loss by damping is compensated by the self--sustaining force. Mathematically, the connection between $\mu_j$ and $f_0$ is expressed by eqs.~(\ref{st2}). In the literature on nonlinear oscillating systems, the limit of weak damping/forcing is also known as the \emph{backbone} approximation \cite{Nayfeh,EPJB}. As in the previous section, the following  results correspond to the case where oscillators $1$ and $2$ have, respectively, hardening and softening nonlinearity ($\beta_2<0<\beta_1$), with $\omega_1<\omega_2$. 

In eqs.~(\ref{st1}), the backbone approximation amounts to taking $f_0 = 0$. Together with the first of eqs.~(\ref{st2}), this yields
\begin{equation} \label{bb1}
\Omega= \frac{k_2 \omega_1 +k_1 \omega_2}{k_1+k_2}, \ \ \ \ \ A_j=\frac{1}{\mu_j \omega_j}\sqrt{\frac{\omega_2-\omega_1}{k_1+k_2}},
\end{equation}
where
\begin{equation}
k_1=\frac{3 \beta_1}{8\mu_1^2 \omega_1^3}, \ \ \ \ \ k_2=-\frac{3 \beta_2}{8 \mu_2^2 \omega_2^3},
\end{equation}
are both positive constants. The second of  eqs.~(\ref{st2}), in turn, implies $\cos \Delta = -1+f_0^{-1} \sqrt{(\omega_2-\omega_1)/(k_1+k_2)}$.   It can be seen that, in the first of eqs.~(\ref{bb1}), the synchronization frequency $\Omega$ reduces to $\Omega_{z=0}$, given by eq.~(\ref{stabfrec}), if the frequency stabilization condition (\ref{cond}) holds. 

Using these results and eq.~(\ref{ec1a}), always within the limit of weak damping, we can evaluate how much $\Omega$ deviates from its stabilized value $\Omega_{z=0}$ if the stabilization condition is not met. We find
\begin{equation}
\Omega=\Omega_{z=0}  + z \frac{\mu_1 \mu_2 a^2}{\mu_1+\mu_2} = \Omega_{z=0}  + z 
\frac{\mu_1 \mu_2}{\mu_1+\mu_2} \frac{\omega_2-\omega_1}{k_1+k_2}.
\end{equation}
This expression makes it possible to compute the deviation $\Omega-\Omega_{z=0}$ in terms of the mechanical parameters of the oscillators and the damping coefficients. The frequency deviation is proportional to the quantity $z$, defined in eq.~(\ref{z}), and to the difference of the natural frequencies, $\omega_2-\omega_1$. As expected, the closer the two frequencies to each other, the more robust frequency stabilization is. 

In order to evaluate the stability of synchronized oscillations in the backbone approximation, we use the fact that the sum of all the eigenvalues of the Jacobian (\ref{mat}) equals its trace, which within the approximation is given by $\mathrm{Tr} ( J) = -(\mu_1+\mu_2) (1+2 \cos \Delta)/2(1+\cos \Delta)$. Moreover, inspection of the characteristic polynomial of $J$ makes it possible to estimate its only real nontrivial root as $\lambda_\mathrm{R} = - \mu_1 \mu_2(k_1+k_2)/2(k_1 \mu_1+k_2 \mu_2)$. Thus, the real part of the complex eigenvalues, whose sign determines the stability of synchronization, is given by $\mathrm{Re}(\lambda_{\rm C}) = [\mathrm{Tr}(J)-\lambda_\mathrm{R}]/2$, or
\begin{equation} \label{long}
\mathrm{Re}(\lambda_{\rm C}) =  
\frac{\mu_1+\mu_2}{2}
\left[ \frac{f_0}{2} \sqrt{\frac{k_1+k_2}{\omega_2-\omega_1}}-1\right] 
+\frac{k_1+k_2}{4} \frac{\mu_1 \mu_2}{k_1 \mu_1+k_2 \mu_2} .
\end{equation}
Note, in particular, the direct dependence of $\mathrm{Re}(\lambda_{\rm C})$ on the combination $f_0/\sqrt{\omega_2-\omega_1}$. In agreement with the cases considered in the previous section, a growth in the coupling force amplitude $f_0$, with all other parameters fixed,  leads to instability of synchronized motion. Equation (\ref{long}) moreover shows that, at least in the weak damping limit, the same effect is obtained when the natural frequencies $\omega_1$ and $\omega_2$ become closer to each other. 

\vspace{10 pt}

{\bf Discussion and conclusion} -- Motivated by potential applications of micromechanical oscillators to miniaturized time--keeping devices, we have analyzed theoretically the joint dynamics of two coupled Duffing oscillators with nonlinearity of opposite signs. In fact, different oscillation modes of clamped--clamped  beams have been experimentally shown to display Duffing (i.e. cubic) nonlinearity of either sign \cite{Nat}. Our main goal has been to find an explicit condition under which the frequency of synchronized motion is independent of the amplitude of the driving force, thus suppressing the undesirable amplitude--frequency effect. Frequency stabilization, which results from the mutual compensation of opposite nonlinear responses, is exact within our analytical multiple scale approach, and holds to a very good approximation in numerical solutions to the equations of motion. Our conclusions should apply to real oscillators as long as the Duffing model correctly describes their dynamics, as quantitatively demonstrated in recent experiments on vibrating micromechanical c--c beams \cite{PRLnew}. The mechanism of frequency stabilization proposed here and those considered in previous work are based on qualitatively different physical principles (nonlinearity compensation vs. internal resonance \cite{Nat,EPJB}). However, on the basis that the Duffing equation provides a good representation of real oscillators of that kind \cite{EPJB,EPJST,PRLnew}, the analytical methods are much the same.

The condition of frequency stabilization, eq.~(\ref{cond}), involves mechanical properties of the oscillators ---namely, the natural frequencies $\omega_j$ and the cubic coefficients $\beta_j$--- as well as the damping coefficients $\mu_j$, which depend not only on mechanical properties but also on the environment where the oscillators are embedded. This suggest that, in an experimental implementation or a technological application to MEMS or NEMS \cite{microos,review}, the stabilization condition could be achieved in two steps: first, a coarse tuning during structural design, which fixes $\omega_j$ and $\beta_j$; second, a finer tuning during operation by controlling $\mu_j$ through, for instance, adjusting the pressure of the atmosphere which surrounds each oscillator. This observation has prompted us to consider one of the damping coefficients ($\mu_2$) as a varying parameter in the presentation of our results (see figs.~\ref{fig2} and \ref{fig3}). 

An unusual feature in the joint dynamics of our coupled Duffing oscillators is that synchronization becomes unstable as the coupling strength grows. Usually, in fact ---and this is also suggested by intuition--- synchronized motion of coupled dynamical systems is increasingly stable as their interaction strengthens \cite{sync,us}. This peculiarity may be understood in terms of the nonlinear response of the oscillation frequency of each individual Duffing oscillator to the growth of the driving force, i.e.~the a--f effect itself, eq.~(\ref{Omega}). In the absence of coupling, the hardening nonlinearity of oscillator $1$ makes that, as the self--sustaining force grows, its frequency increases from the natural frequency $\omega_1$ upwards. Conversely, because of its softening nonlinearity, the frequency of oscillator $2$ decreases, from $\omega_2$ downwards. Since $\omega_1 < \omega_2$, the respective oscillation frequencies first approach each other as the forcing grows but, above a certain force value at which the two frequencies coincide, they become progressively more separated. For sufficiently strong self--sustaining forces, consequently, the frequencies at which the two oscillators effectively vibrate can be very different from each other, which inhibits their synchronization when they are coupled.  In fact, large differences between  individual frequencies is a well--known destabilizing factor for mutual synchronization under a broad class of coupling  schemes  \cite{sync,us}.  In our system, the coupling force plays, at the same time, the role of self--sustaining force. According to the above discussion, thus, strengthening the coupling brings about a separation of the individual frequencies. Synchronization becomes increasingly difficult and, as shown by our results, is eventually  destabilized.

Similarly unusual is the fact that destabilization of synchronized motion is favored by the proximity of the natural frequencies $\omega_1$ and $\omega_2$, as shown by eq.~(\ref{long}). An explanation for this observation may reside in the fact that, according to the second of eqs.~(\ref{bb1}), the oscillation amplitudes tend to vanish as the natural frequencies approach each other. Such small amplitudes may not be able to sustain the rate of energy dissipation necessary to counteract the energy input coming from the self--sustaining force, thus requiring the system to perform larger, but not coherent, oscillations. 

A natural continuation of the present contribution would be to set up an experiment, either with micromechanical c--c beams or with another realization of the Duffing oscillator, to verify whether frequency stabilization can be established as proposed here. Our results provide a starting point for the quantitatively detailed design of such study. 

\vspace{10 pt}

{\bf Acknowledgements} -- Collaboration with D.~Antonio, S.~Arroyo, Changyao Chen, D.~Czaplewski,  J.~Guest, D.~L\'opez,  and F.~Mangussi, as well as financial support from ANPCyT, Argentina (PICT 2014--1611), are gratefully acknowledged.

\end{document}